# Resource-Aware Network Topology Management Framework

**Aaqif Afzaal Abbasi[1], Shahaboddin Shamshirband[2,3*], Mohammed A. A. Al-qaness[4], Almas Abbasi[5], Nashat T. AL-Jallad[6,7], Amir Mosavi[8,9,10,11]**

[1]Department of Software Engineering, Foundation University, Islamabad, 44000 Pakistan
aaqif.afzaal@fui.edu.pk

[2]Department for Management of Science and Technology Development, Ton Duc Thang University, Ho Chi Minh 758307, Vietnam
[3]Faculty of Information Technology, Ton Duc Thang University, Ho Chi Minh 758307, Vietnam
shahaboddin.shamshirband@tdtu.edu.vn

[4]School of Computer Science, Wuhan University, Bayi Road No. 299, Wuhan 430072, China
alqaness@whu.edu.cn

[5]Department of Computer Science, International Islamic University, Islamabad 44000, Pakistan
almas.abbasi@iiu.edu.pk

[6]School of Computer Science and Technology, Wuhan University of Technology, Wuhan 430079, China
[7]School of Computer Science, Palestine Technical University, Tulkarem 44864, Palestine
jallad@whut.edu.cn

[8]School of the Built Environment, Oxford Brookes University, No.5, Jack Straws Lane, Oxford OX3 0BP, UK.
a.mosavi@brookes.ac.uk

[9]Kalman Kando Faculty of Electrical Engineering, Obuda University, Becsi ut 94-96, Budapest 1034, Hungary.
amir.mosavi@kvk.uni-obuda.hu

[10]Institute of Structural Mechanics, Bauhaus University Weimar, Marienstraße 15, D-99423 Weimar, Germany
amir.mosavi@uni-weimar.de

*Abstract: Cloud infrastructure provides computing services where computing resources can be adjusted on-demand. However, the adoption of cloud infrastructures brings concerns like reliance on the service provider network, reliability, compliance for service level agreements (SLAs), etc. Software-defined networking (SDN) is a networking concept that suggests the segregation of a network's data plane from the control plane. This concept improves networking behavior. In this paper, we present an SDN-enabled resource-aware topology framework. The proposed framework employs SLA compliance, Path Computation Element (PCE) and shares fair loading to achieve better topology features. We also present an evaluation, showcasing the potential of our framework.*

# 1 Introduction

Cloud computing technology provides computing and networking services over the internet [1]. Computational and I/O resource management in cloud computing is a challenging task. Different methods have been adopted to address the computation and I/O management challenges in cloud systems. Therefore, the success of any cloud management software depends on the efficiency of the system through which it can utilize the underlying networking resources [2,3,4]. Figure 1 shows a generic view of services provided by a conventional cloud resource management system. It includes a set of virtual machines (VMs) operating on a network operating system (hypervisor). A VM is an emulation of a physical computing machine. It provides the functions of a computer system by using the resources of the underlying hardware and software resources. User applications are hosted as applications (APPs) on guest operating systems that operate on these VMs. For a detailed study of the topology and virtualization system technology, literature research is available in [1,3].

There have been many attempts to make networks more manageable and secure. Various methods have been adopted to deliver resource management features in a cloud environment. However, one of the drawbacks in cloud services delivery is that consumers are kept unaware of the details of how cloud services and features are provided. In effect, users focus on what really matters to them, i.e. consuming a service. Similarly, the cloud service providers focus only on aspects of their domain that are largely nontransparent to the end consumers.

In software-defined networking (SDN) [5], the control plane of a network element is separated from its data plane functions. SDN technology is used in data centers to effectively manage network traffic. The SDN principles can also be applied to other areas such as storage, security and service level agreement. Software-defined cloud computing (SDCC) in this term in an approach where all aspects of a data center providing services to the users are software-defined [6]. The principles and concepts of SDCC provide an easy way for reconfiguration and adaptation of physical resources to adjust QoS demands [7,8]. Figure 2 shows the architecture of an SDN enabled cloud resource management system.

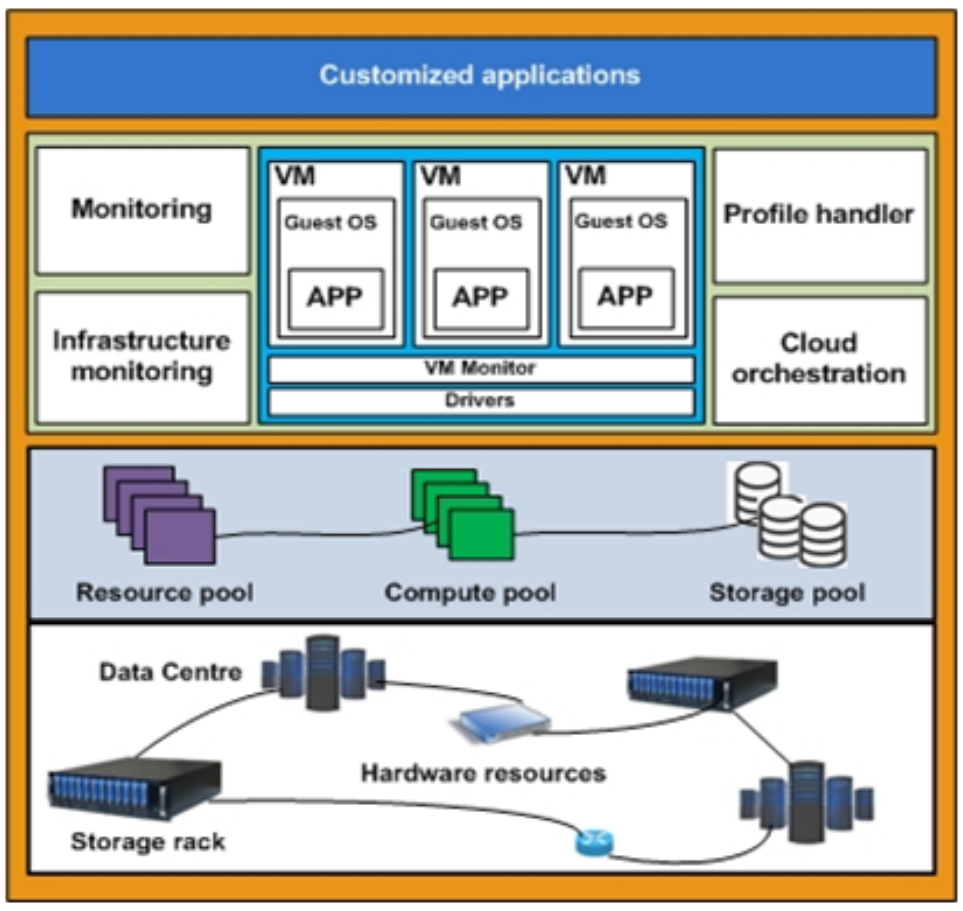

Figure 1
Cloud resource management infrastructure

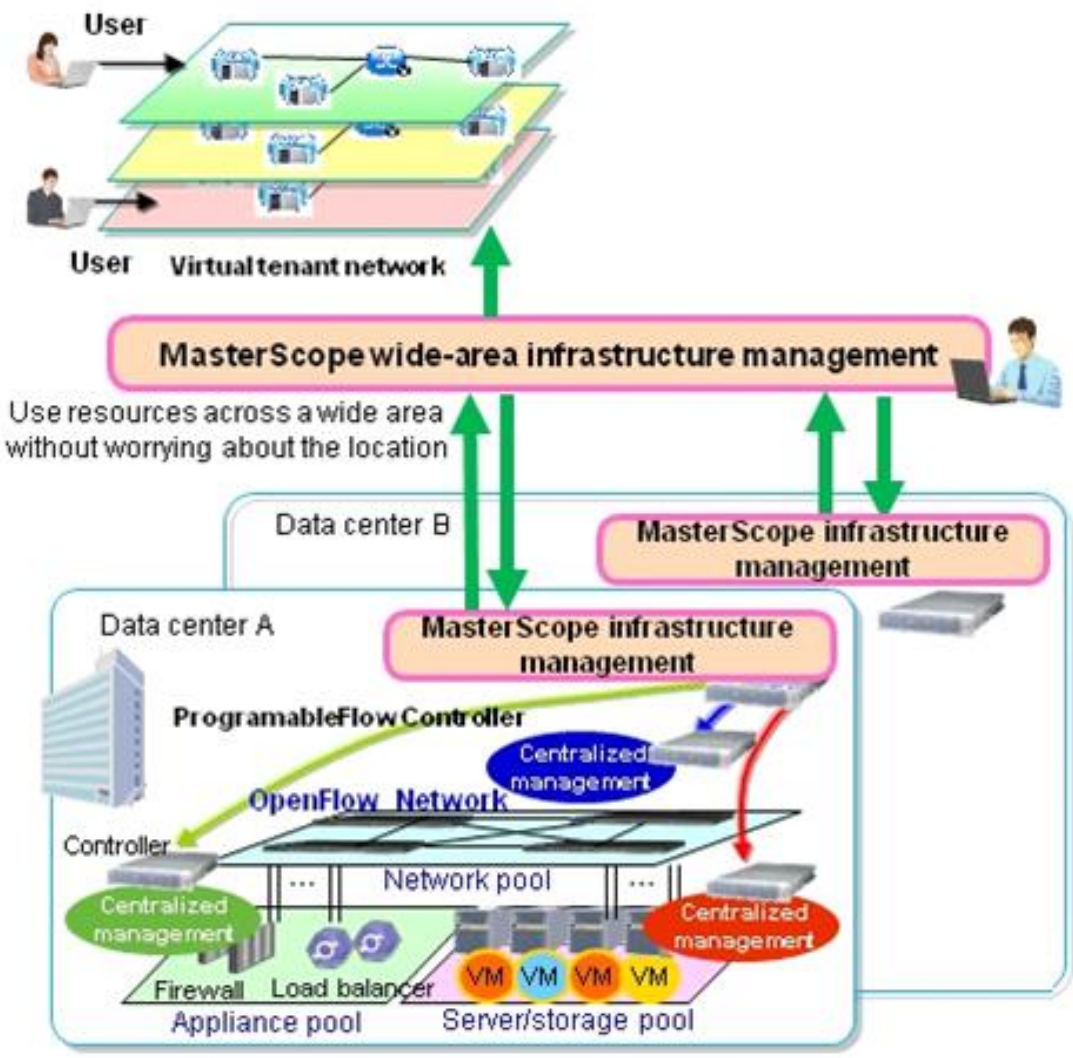

Figure 2
SDN-enabled cloud resource management

In this paper, a network topology management framework is presented. We explain the structure of its topology administration features. The paper also includes a discussion on the architectural developments made in traditional and SDN enabled cloud resource management systems. A thorough analysis of topology management functions has also been presented in section 4. Finally, we present the results of our framework.

The paper is organized as follows. Section 2 describes the related work. Section 3 discusses the architecture and working of the presented framework. Section 4 presents its working and Section 5 presents the evaluation and analysis of results. Finally, Section 6 concludes the paper.

# 2. Related Work

## 2.1 Cloud resource management issues

A gossip-based protocol is presented in [9]. The protocol uses a heuristic solution to solve resource allocation challenges. By delivering a simulation-based approach to solve this problem, the protocol can be employed to achieve fairness in resource allocation scenarios. The issues pertaining to network utility maximization (NUM) in SDN controllers can be addressed by using variable resource decomposition methods which can also look after the update rules in parallel.

The SLA based resource allocation challenges have been discussed in [10]. Due to the recent emergence of complex IT technologies the user applications are becoming complicated. In particular, the cloud management framework in [11] discusses the resource provisioning in a cloud environment. Among the multiple issues faced by cloud services, the most persistent problem is that users are unaware of the services provisioning methodology. The recent surveys on the topic indicate that cloud system developers should develop tools to automate cloud operator tasks. This will improve cloud services delivery and will bring transparency to the technology audit mechanism. Resource management strategies for improving network overheads are discussed in [12]. The research analyzes the pros and cons of these strategies particularly in terms of performance costs and services stability. Other techniques used for controlling and managing network traffic across WAN includes the use of multiprotocol label switching (MPLS) over SDN managed carrier connections that can handle incoming network traffic from multiple locations.

In cloud-based systems, multiple user services are entertained simultaneously. This is possible due to the recent advancements in efficient parallel data processing techniques. Efficient parallel data processing is described in [13]. It presents Nephele, a framework that uses the benefits of resource provisioning services offered by IaaS clouds. The on-demand service provisioning framework for grid computing is presented in [14] where the system allocates resources to users on the basis of their profile and service usage. A profile-based approach is presented in [15] where the user profile is used to evaluate resource usage. A service optimization framework for risk-aware resource provisioning of dynamic resource allocation is presented in [16] where the workload of multiple clients is evaluated under the uncertainty of workloads. SDNs are extensively used for risk assessment to redefine network operations at runtime. This is due to their resilience when used as a control parameter to administer the underlying hardware infrastructure.

## 2.2 SDN-based cloud resource management

The software-defined cloud functions of a data center are administered by an open-access user interface. This helps in discouraging proprietary software from handling network resource management functions. A software-defined resource manager automatically manages network data, offering easier administration. It can work with existing resource management solutions allowing applications to share common resource management platforms.

Harmony [17] is proposed to manage various aspects of software-defined clouds. It reduces workload dependencies between different tasks. In order to achieve fault tolerance, a model framework has been presented in [18] which realize the true benefits of SDNs in data centers. SDN-based orchestration technologies coordinate together to provide balanced composite cloud and network services. These technologies also ensure that VM allocations in the network topologies are based on estimations of switch/link and server loads.

## 2.3 SDN as enabling technology to administer resource management

SDNs bring network awareness to network control features. SDN controllers can read the entire network topology through subnets. Subnets use available network resources to constitute a logical topology within a network. In [19], a software-defined interface is presented which uses pluggable modules for scheduling and fault management of a network. This enables SDN applications to deploy network control functionalities in a practical multi-tier cloud infrastructure as shown in Figure 3. A software-defined resource manager discovers the inventory of links in a given network by plotting all possible paths across the network. Therefore, if all the applications use the best available path strategy for performing network functions, it can result in bringing greater resource administration and agility. Research studies conducted in [4] are aimed at reducing network operation costs by either combination with virtualization of network services through the use of SDN-enabled resource allocation techniques. Network operation and resource allocation through centralized data streams also benefit network users in simplifying network software upgrades.

# 3. System Architecture

The proposed framework consists of the SDN application programming interface (API), a cloud resource infrastructure and underlying computational resources as shown in Figure 3. The application management APIs consist of the cloud management console, topology manager and admission controller. The SDN-based API manages topology functions in a cloud environment. Cloud management console acts as an interface between the user and applications.

The SDN API lies next to a network of underlying cloud resources, which logically control resource management operations. It is followed by a layer of virtual and physical components. The physical part typically includes the servers, storage media, and network peripherals. The virtual layer consists of the software deployed across the physical layer. The arrangement of the network entities is similar to [20]. Below we provide a short description of each component.

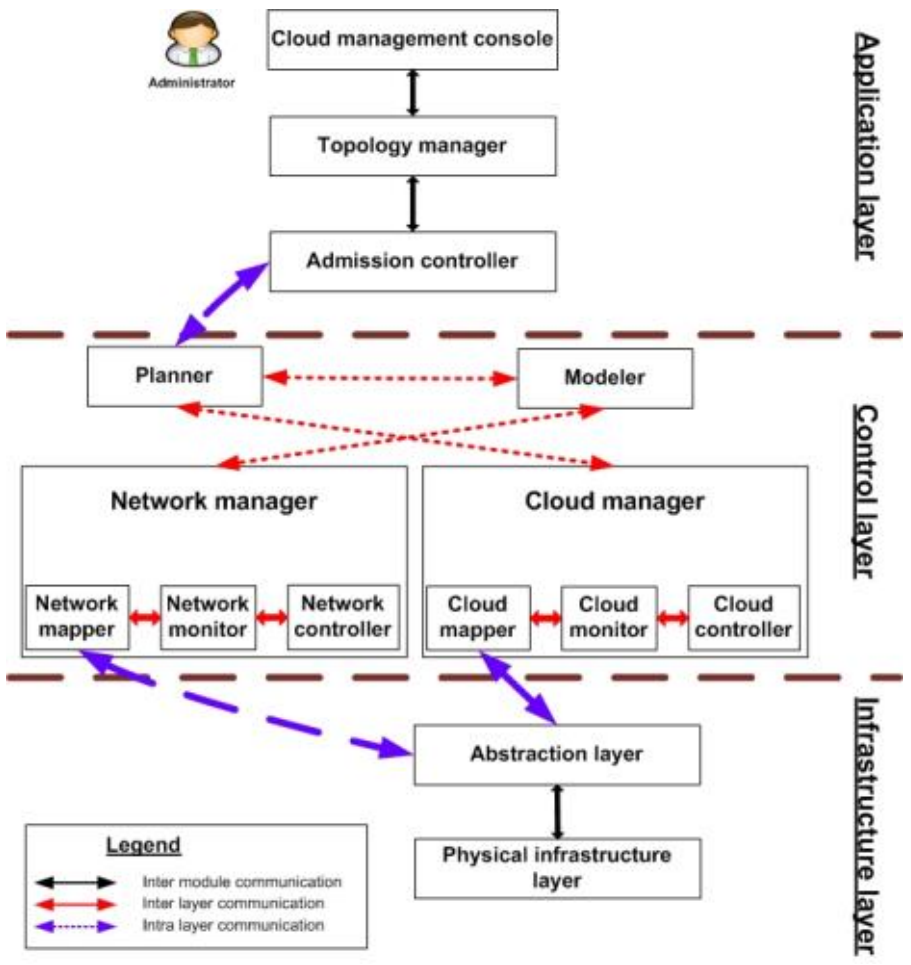

Figure 3

SDN-enabled cloud resource management

**Cloud management console**: The console facilitates cloud resource management functions to express high-level policies on the underlying network infrastructure.

**Topology manager**: It maintains a directory of overall network mappings. It enables access to a seemingly infinite pool of computing resources using open federated cloud computing architecture like RESERVOIR [21]. This facilitates the administrator in gaining an understanding of available network resources through a service provider network.

**Admission controller**: The admission controller receives and evaluates the user requests on the basis of prescribed factors that are defined through SLAs.

**Planner**: A planner in consultation with the modeler, network manager, and cloud manager determines the location of hosting the services.

**Modeler**: The modeler compares the planner request with the network manager and cloud manager to plot a resource utilization scope that can be sent to the topology manager for updating the network directory.

**Cloud and network manager**: Cloud and network managers consolidate data center infrastructure resources. This consolidation is performed at both physical and logical levels. They monitor network status and manage both physical and virtual infrastructures.

**Abstraction layer**: The layer provides an abstraction of logical deployment of physical hardware for all devices in the cloud infrastructure.

**Physical infrastructure layer**: It consists of the physical resources of a data center network including computing devices, storage devices, servers, and other networking equipment.

The proposed topology manager provides an insight into the hierarchical structure and state of the cloud. By using SDN's flexible nature, the topology manager helps in organizing software and hardware into zones, blocks, servers, nodes, resource pools, and software deployments. In the next section, we explain the working of the topology manager in detail.

# 4. Topology Management Framework

## 4.1 Topology management

Route analytics and SDN together improve the availability of network routes and traces in data centers. SDNs allow network administrators to manage real-time network-wide abstractions into topology control. This helps the SDN-enabled data center infrastructure to use the latest traffic status and workload profiles. In this regard, TOSCA is presented in [22]. It is an industrially-endorsed standardization for topology specifications. By using SDN control functions, it creates a blend of network traffic requirements. This helps network and cloud applications to derive the relationship between a service and its behavior in the network. A topology manager's implementation provides a complete view of the availability of devices and resources. Its function is independent of the vendor-specific monitoring technologies. The proposed topology manager uses a combination of techniques for monitoring and customization of third party services. Below we give a short description of its components and their functions.

**Application handler**: When topology requests are scheduled for deployment, the application handler registers these services. It performs multiple checks on topology requests and later sends them for the compatibility checks.

**Service handler**: Once a topology request confirms SLA requirements, it is accessed by a service handler that compares its requirements with available virtual infrastructure.
**Path Computation**: PCE [23] helps in reducing resource-based computational constraints during path computation by considering multiple constraints. In our model, we first ensure that the experimental test-bed has sufficient resources for task execution. Then, we use PCE for path computation.

The topology management operation initiates when an application requests for grant of resources. A graphical portrayal of the topology manager scheme is given in Figure 4. The application handler performs a check operation on SLA violations. If SLA violations are made the process is terminated. If not, a check is performed for available resources. The applications with limited resources are dropped. All dropped processes are prompted to the application handler for rejection. Those processes with enough resources are forwarded to the PCE module. The PCE module is used to define the suitable path between the traffic source and destination. It then assigns optimal resources to the process/ request. The use of PCE helps in reducing the processing overheads as it uses the previously calculated paths from the path allocation table.

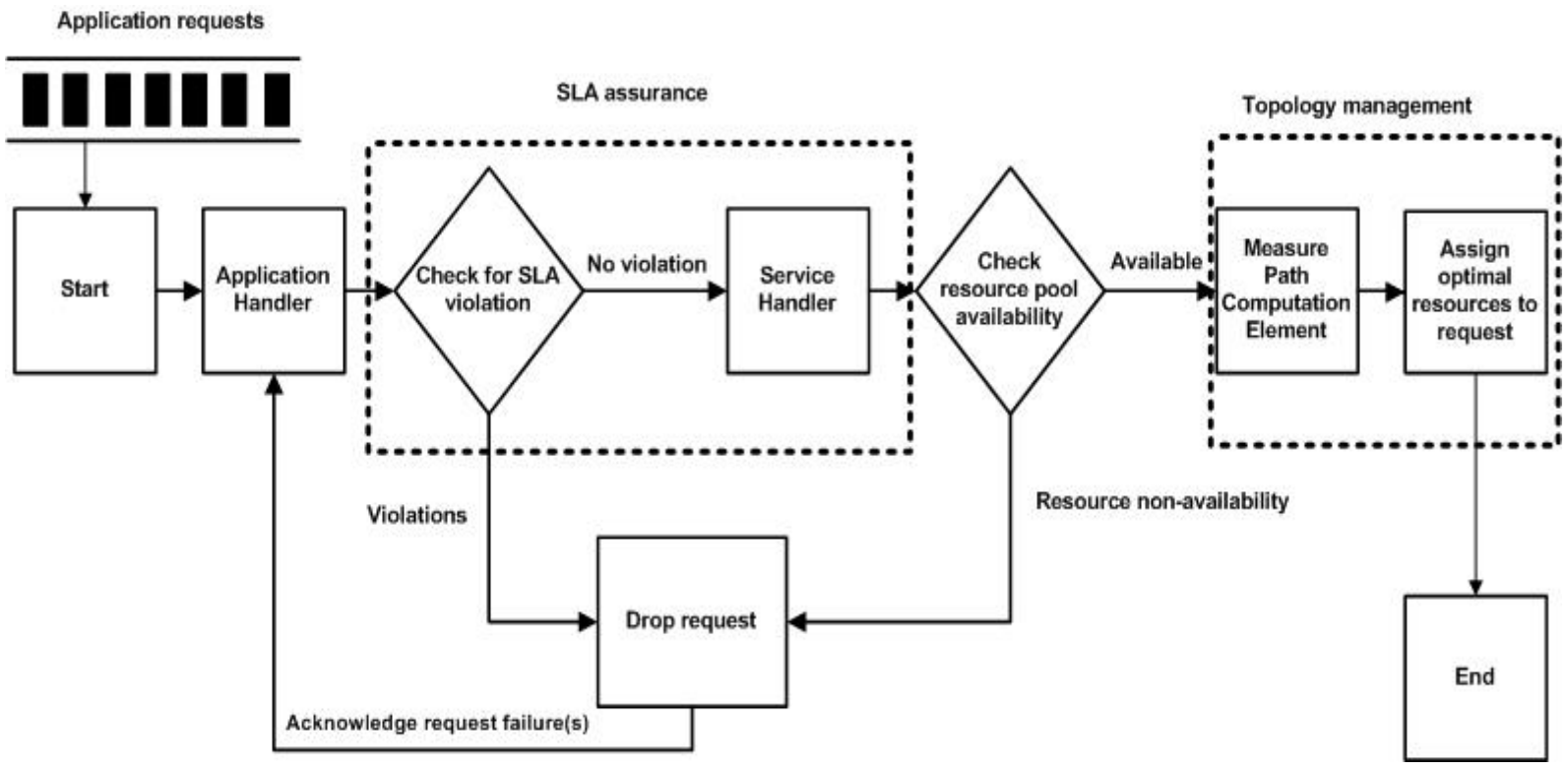

Figure 4
Topology management - Process flow diagram

We evaluate the proposed cloud management console prototype on an Intel Xeon CPU E5-2650 v2 with a clock speed of 2.60GHz and 8 cores (16 threads). It is developed in Python version 2.7.9. By using Ubuntu 12.04 environment, we use two 8 slot SDN enabled HP 12508E switches, having a 10.8 Tbps maximum switching capacity. We measure the framework for throughput and CPU resource utilization efficiency.

Our results (Figure 6) show that the proposed framework can handle an increasing number of requests. This is because the allocation scheme used in share fair systems is not the same as utilization. Normally a request allocating 50-60 percent of CPU resources to process an application, might only use a part of these resources. Regardless of how big resource allocation is, a resource requesting query always receives 100 percent of the processing capacity. Maintaining a balance among resource usage and allocation is complicated sometimes. Therefore, the allocation of either a small or big chunk of CPU share to a busy workload might not solve the problem, yet it may result in slower performance.

We conducted a performance evaluation of our proposed framework by submitting admission control requests and then measuring the system's CPU and memory utilization. We used conditional statements to ensure that SLA conditions are not violated for the topology management scheme. In our experiments, we used the Batch workload. By submitting the Batch workload jobs through a job queue, it was ensured that the submitted data load runs unconstrained, and is free from bottlenecks.

Figures 5 and 6 show the CPU and memory resources utilization. We compared the proposed framework efficiency with Realistic and Capacity-aware admission control schemes for 9 instances. The realistic approach employs product logic for modeling requests. A detailed study of product logic is presented in [24]. The capacity-aware admission control scheme uses optimized risk values for CPU and I/O functions and employs real-time values for memory mapping of received admission control requests. From the results can be seen that Realistic and Capacity aware schemes are constraining network resources. These techniques achieved better performance for one capacity (CPU usage) at the expense of others (memory usage). This results in presenting a more asymmetric behavior. However, the results show small signs of jitter due to the continuous increase in the amount of resource utilization. In Figure 7, we present the overall resource utilization comparison. Results demonstrate that by using our proposed methodology, resource utilization is even lower than the average of Realistic and Capacity aware schemes. Due to these schemes, we can confirm the effectiveness of our proposed framework.

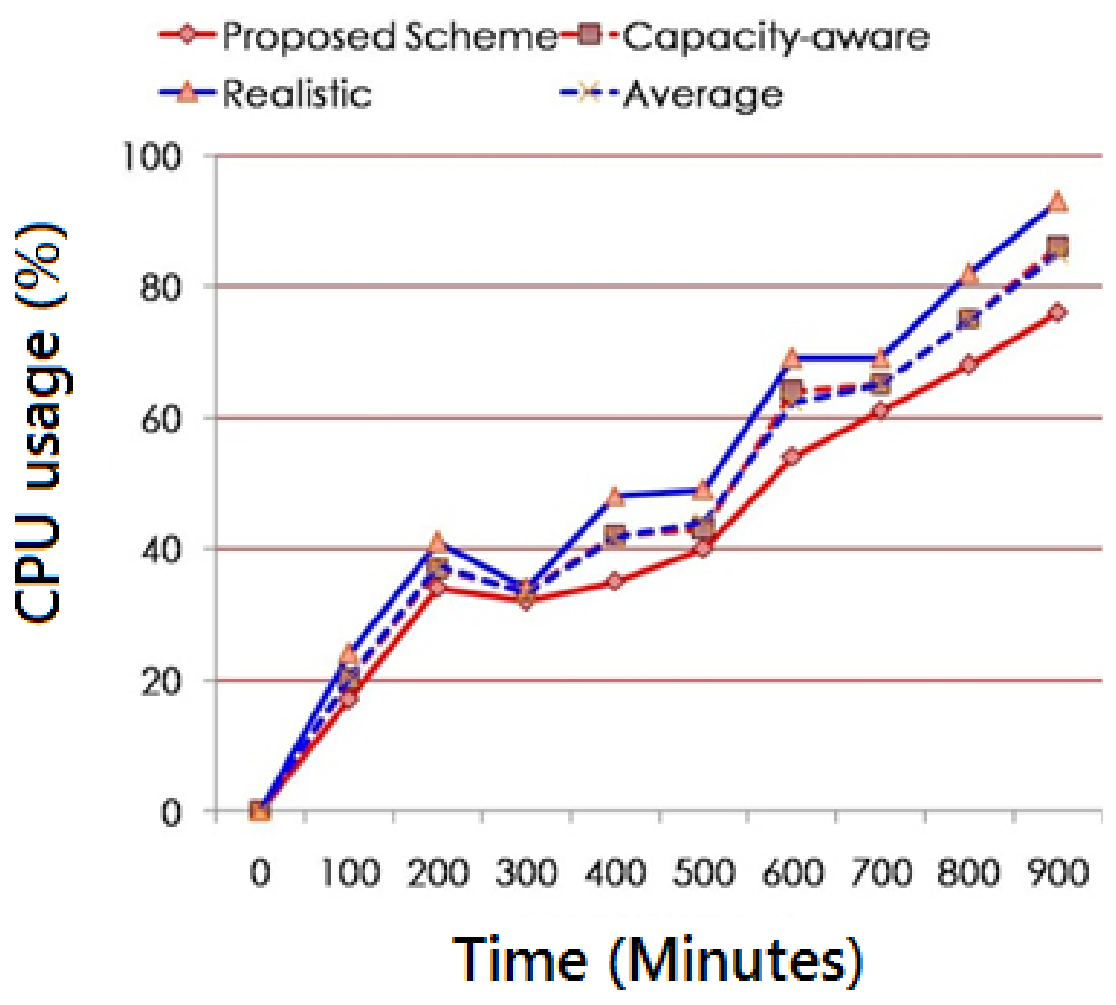

Figure 5
CPU resources utilization

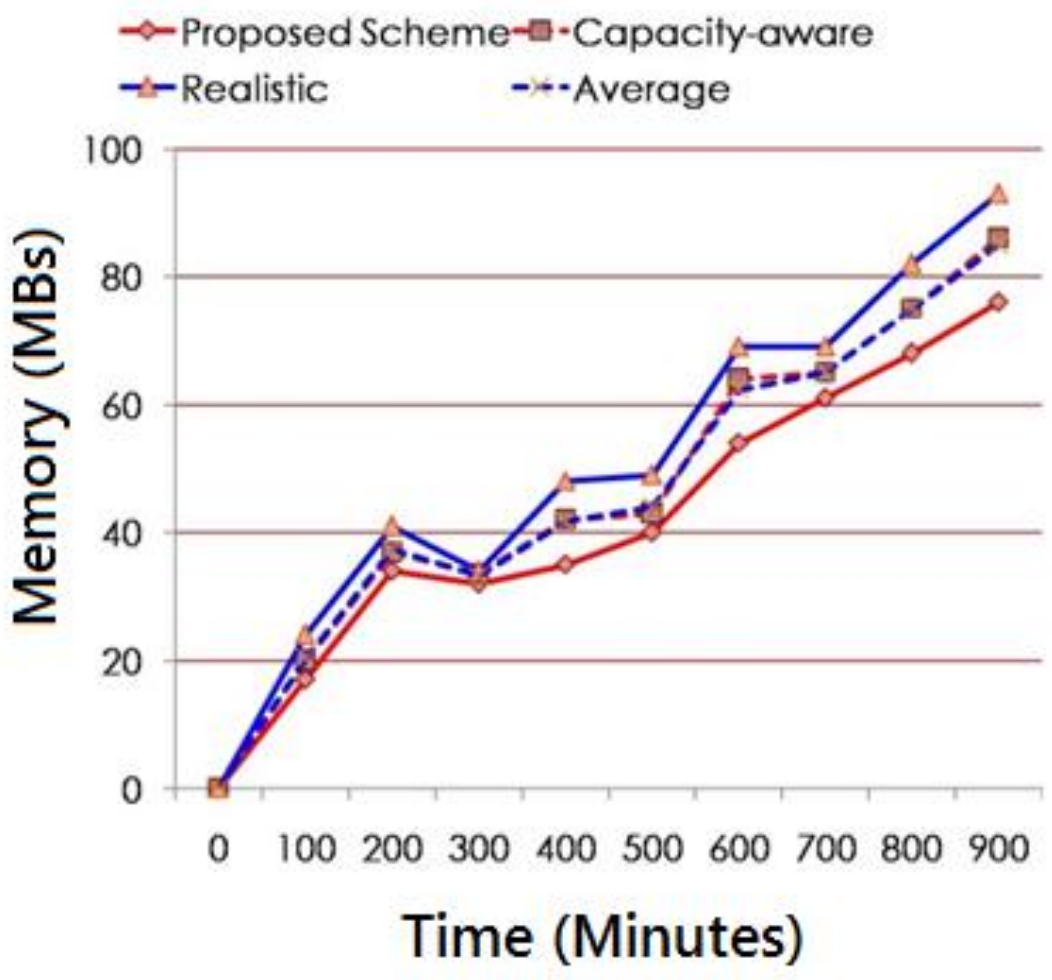

Figure 6
Memory resources utilization

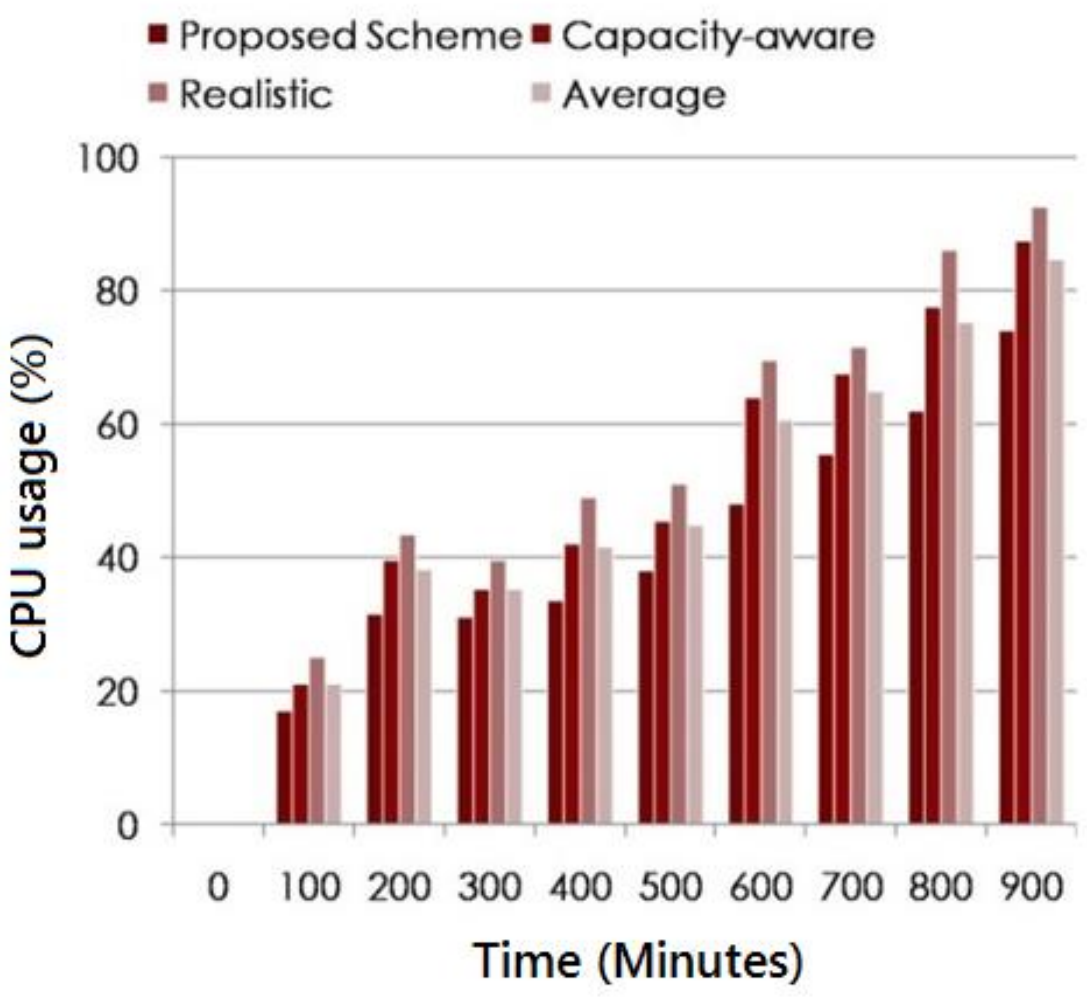

Fig. 7
Overall resource utilization comparison

# Conclusions

SDN concepts bring great potential and efficiency to reduce the complexity of network control. To efficiently manage and configure the network, it needs to have up-to-date information about the state of the network, in particular, its topology.

The paper presents a network topology framework. It utilizes SDN–enabled infrastructure for improving topology management functions. The proposed framework employs SLAs, PCE and shares fair loading as a means for improving resource administration. The framework facilitates in reducing complexities for resource allocation problems. We evaluate our system on limited real-time controlled data center traffic. We believe that the presented work provides a foundation for developing a more efficient topology management infrastructure. In the future, we plan to improve our system's planner design so that it can effectively handle VM placement and allocation related challenges. We also plan to reduce VM overheads by improving the topology discovery feature in a data center environment.

# Acknowledgement

We acknowledge the financial support of this work by the Hungarian State and the European Union under the EFOP-3.6.1-16-2016-00010 project and the 2017-1.3.1-VKE-2017-00025 project.

## References

[1] Jin, Hai, Aaqif Afzaal Abbasi, and Song Wu. "Pathfinder: Application-aware distributed path computation in clouds." *International Journal of Parallel Programming* 45.6 (2017): 1273-1284.

[2] Abbasi, Aaqif Afzaal, and Hai Jin. "v-Mapper: An Application-Aware Resource Consolidation Scheme for Cloud Data Centers." *Future Internet* 10.9 (2018): 90.

[3] Gulati, Ajay, Arif Merchant, and Peter J. Varman. "mClock: handling throughput variability for hypervisor IO scheduling." Proceedings of the 9th USENIX conference on Operating systems design and implementation. USENIX Association, 2010.

[4] Mitzenmacher, Michael. "The power of two choices in randomized load balancing." *IEEE Transactions on Parallel and Distributed Systems* 12.10 (2001): 1094-1104.

[5] Kim, Hyojoon, and Nick Feamster. "Improving network management with software defined networking." *IEEE Communications Magazine* 51.2 (2013): 114-119.

[6] Abbasi, Aaqif Afzaal, Hai Jin, and Song Wu. "A software-defined cloud resource management framework." Asia-Pacific Services Computing Conference, Springer, 2015.

[7] Fundation, Open Networking. "Software-defined networking: The new norm for networks." ONF White Paper 2 (2012): 2-6.

[8] Nunes, Bruno Astuto A., et al. "A survey of software-defined networking: Past, present, and future of programmable networks." *IEEE Communications Surveys & Tutorials* 16.3 (2014): 1617-1634.

[9] Wuhib, Fetahi, Rolf Stadler, and Mike Spreitzer. "Gossip-based resource management for cloud environments." IEEE International Conference on Network and Service Management (CNSM), 2010.

[10] Buyya, Rajkumar, et al. "Cloud computing and emerging IT platforms: Vision, hype, and reality for delivering computing as the 5th utility." *Future Generation computer systems* 25.6 (2009): 599-616.

[11] Cunningham, S. J. "Developing innovative applications of machine learning." Proc. Southeast Asia Regional Computer Confederation Conference. 1999.

[12] Juve, Gideon, and Ewa Deelman. "Resource provisioning options for large-scale scientific workflows." IEEE Fourth International Conference on eScience, 2008.

[13] Warneke, Daniel, and Odej Kao. "Exploiting dynamic resource allocation for efficient parallel data processing in the cloud." *IEEE Transactions on Parallel and Distributed Systems* 22.6 (2011): 985-997.

[14] Leivadeas, Aris, Chrysa Papagianni, and Symeon Papavassiliou. "Efficient resource mapping framework over networked clouds via iterated local search-based request partitioning." *IEEE Transactions on Parallel and Distributed Systems* 24.6 (2013): 1077-1086.

[15] Yang, Jie, Jie Qiu, and Ying Li. "A profile-based approach to just-in-time scalability for cloud applications." IEEE International Conference on Cloud Computing, 2009.

[16] Kusic, Dara, and Nagarajan Kandasamy. "Risk-aware limited lookahead control for dynamic resource provisioning in enterprise computing systems." *Cluster Computing* 10.4 (2007): 395-408.

[17] Grandl, Robert, et al. "Harmony: Coordinating network, compute, and storage in software-defined clouds." ACM Proceedings of the 4th annual Symposium on Cloud Computing, 2013.

[18] Baset, Salman A., et al. "Toward achieving operational excellence in a cloud." *IBM Journal of Research and Development* 58.2/3 (2014): 4-1.

[19] Lin, Thomas, et al. "Enabling SDN applications on software-defined infrastructure." IEEE Network Operations and Management Symposium (NOMS),2014.

[20] Mell, Peter, and Tim Grance. "The NIST definition of cloud computing." (2011).

[21] Rochwerger, Benny, et al. "The reservoir model and architecture for open federated cloud computing." *IBM Journal of Research and Development* 53.4 (2009): 4-1.

[22] Mousavi, S., Mosavi, A., Varkonyi-Koczy, A.R. and Fazekas, G., 2017. Dynamic resource allocation in cloud computing. Acta Polytechnica Hungarica, 14(4), pp.83-104.

[23] Farrel, Adrian, J-P. Vasseur, and Jerry Ash. A path computation element (PCE)-based architecture. No. RFC 4655. 2006.

[24] Vázquez, Carlos, et al. "A fuzzy approach to cloud admission control for safe overbooking." International Workshop on Fuzzy Logic and Applications. Springer, 2013.